\documentclass{IET-Conf-Paper}
\usepackage{graphicx}
\usepackage{subcaption}
\usepackage{placeins}

\begin{document}

\title{Sizing of a grid-forming power converter to improve the small-signal stability of an LCC-HVDC system connected to a weak grid}

\author{Anup Joshi\ad{1,2}\corr, Javier Renedo\ad{2}, Xavier Guillaud \ad{1}}

\address{\add{1}{Univ. Lille, Arts et Metiers Institute of Technology, Centrale Lille, Junia, ULR 2697 - L2EP, F-59000 Lille, France}
\add{2}{Red Eléctrica - Redeia, 28108 Alcobendas, Madrid, Spain}
\email{anup.joshi@ree.es, anup.joshi@centralelille.fr}}


\begin{abstract}
Line-commutated converter high-voltage direct current (LCC-HVDC) has proven to be a reliable technology for bulk power transmission over long distances. However, the growing penetration of converter interfaced generation (CIG) is resulting in weaker AC grids, rendering the operation of LCC-HVDC systems vulnerable and posing a serious challenge to their stability. Grid-forming (GFM) controlled voltage source converter (VSC) have been shown to provide stabilizing impact in weak grid conditions. However, the impact of GFM controlled VSCs (GFM-VSC) on stability of LCC-HVDC in weak grid conditions has not been studied in depth in the literature. In this paper, a simplified model of LCC-HVDC is proposed and validated. Then a small-signal state-space model of a system consisting of aforementioned LCC-HVDC, a GFM-VSC and an infinite grid is developed to study the interactions between different components. The small-signal stability analysis shows the stabilizing effect of the GFM-VSC on the stability of the LCC-HVDC link in weak grid condition. Furthermore, the study on the sizing of the GFM power converter reveals that even a modest share of the capacity of the GFM power converter relative to the total nominal apparent power (sum of nominal power of LCC-HVDC and the nominal apparent power of GFM-VSC) is sufficient to ensure the stability of the system, in the test system analyzed in this study. This work just focuses in small-signal stability, but it is important to highlight that other stability phenomena should also be taken into account when selecting the final size of the GFM-VSC.

$ $
\\
This is an unabridged pre-print of the following paper (accepted for publication):
\begin{itemize}
\item A. Joshi, J. Renedo and X. Guillaud, \emph{Sizing of a grid-forming power converter to improve the small-signal stability of an LCC-HVDC system connected to a weak grid,} Proc. 22nd IET International Conference on AC and DC Power Transmission (ACDC Europe 2026), Berlin, Germany, 28-29 April, 2026, pp. 1-6.
\end{itemize}

\end{abstract}

\maketitle

\section{Introduction}

Over the past decades, line-commutated converter high-voltage direct current (LCC-HVDC) has proven to be a robust and reliable technology for massive power transmission over long distances. Despite its undeniable benefits, its operation poses inherent challenges: It requires a stable AC voltage to operate  thus making it vulnerable in case of operation in weak grids \cite{b1}. This issue is exacerbated as the penetration of  converter interfaced generation(CIG) increases, leading to a weaker grid. In recent years, grid forming (GFM) controlled voltage source converters (VSC) with their 'voltage source' behaviour have been touted as promising technology to provide stabilizing capabilities in weak grid condition with high amount of CIGs \cite{b2}. \color{black} It is crucial to assess whether the GFM power converters can enhance the stable operation of LCC-HVDC in weak grid conditions to ensure their operational viability in such specific application scenarios.

\color{black}Early modeling approaches for small signal stability study of LCC-HVDC are based on the analytical framework to study the HVDC-HVAC interaction. In paper \cite{b3}, an analytical model of LCC-HVDC is provided to study the HVDC-HVAC interactions in weak grid scenario by dividing the system into various subsystems such as AC grid, converters and DC grid. In reference \cite{b4}, the authors propose to simplify network parameters not in the vicinity of the dynamic components to be represented by constant admittances in order to reduce the computational burden. The study in \cite{b5} proposes an alternative modeling approach in which the dynamics of LCC-HVDC is mapped onto a mass-damping-spring form. More recent work has expanded to include the multi-infeed LCC-HVDC system \cite{b6} and multi-device systems, including wind turbine generators \cite{b7} to study the interaction between different components.

Beyond modeling, several studies have focused on control aspects and their implication on stability. Reference \cite{b3} provides an insight on the influence of phase-locked loop (PLL) gains while the work in \cite{b8} compares the impact of constant dc voltage control and extinction angle control at the inverter and \cite{b7} presents the impact of both controller parameter and PLL gains on the sub-synchronous oscillation study in a multi-device setup.

Recently, some studies have addressed the issue of vulnerability of LCC-HVDC in weak grids and the positive impact of GFM controlled VSC. Reference \cite{b9} considers a system with multi-infeed LCC-HVDC and modular multilevel converter (MMC) based MMC-HVDC. The stability enhancment from a GFM controlled MMC is documented with the help of transient studies in case of fault and generation loss. Similarly, the study in \cite{b10} shows the ability of a GFM static compensator (STATCOM) in  suppressing the commutation failure in LCC-HVDC in weak grid condition. In paper \cite{b11} the stabilizing capabilities of GFM-VSC in damping the oscillatory modes arising in a multi-converter setup in weak grid conditions is studied.

\color{black}The above studies provide valuable insights into analytical modeling of LCC-HVDC in single and multi-device setup, control system implications and positive impact of GFM-VSC power converters. However, a specific study of sizing a GFM power converter to ensure small-signal stability in a power system consisting of an LCC-HVDC link connected to a weak grid has not been addressed before in the literature. \color{black} This paper combines the elements of the aforementioned studies and provides a methodology to study the interaction between a GFM-VSC, an LCC-HVDC and a weak AC grid. A test system is developed consisting of a GFM-VSC, an LCC-HVDC link and a thevenin equivalent grid. A small-signal model of this test system is developed to study the interaction and to verify the stabilizing effect of the GFM-VSC. Finally, a study is performed for sizing the GFM power converter for the stable operation of LCC-HVDC link in weak grid condition, just focusing on small-signal stability and interactions. Analysis of other phenomena, such as commutation failure or voltage stability, for example, are out of the scope of this paper. The accuracy of the small-signal state-space models have been verified by comparing with the Electromagnetic Transient (EMT) simulation models in MATLAB/Simulink. 

\section{Modeling of LCC-HVDC link and GFM power converter}
The focus of this study is the inverter side converter. As the rectifier side adopts DC current control, the rectifier side and its AC grid can be equivalently represented by a constant DC current source. This approach allows to focus on the study of interaction between the inverter and the AC grid to which it is connected while reducing the modeling complexity. The inverter consists of a 12-pulse LCC and the converter transformer considered is an ideal three-winding Y-$\Delta$ transformer. \color{black} A constant reactive power compensation of 600 Mvar is provided through capacitor banks and filters for a stable LCC operation at normal conditions. \color{black}This simplified scheme of the LCC-HVDC link is shown in Fig. \ref{fig:switch} and hereafter referred to as the switching model. An average model of the above switching model is then developed on which the small-signal state-space model of the LCC-HVDC link is based.

\subsection{Simplified non-linear model of LCC-HVDC link}
\begin{figure}[h]
  \centering
  \begin{subfigure}[b]{1\columnwidth}
    \centering
    \includegraphics[width=0.9\linewidth]{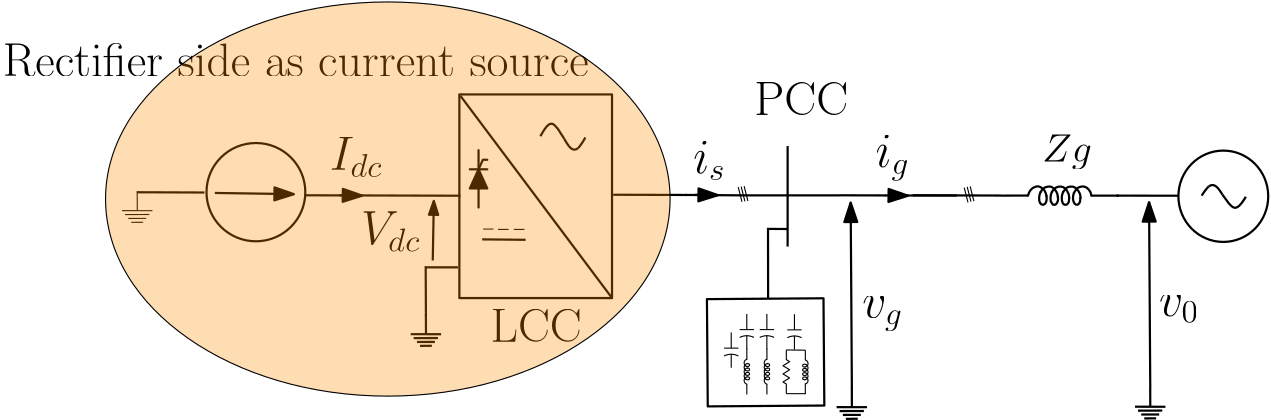}
    \caption{Switching model}
    \label{fig:switch}
  \end{subfigure}
  \vspace{0.5em} 
  \begin{subfigure}[b]{1\columnwidth}
    \centering
    \includegraphics[width= 0.9\linewidth]{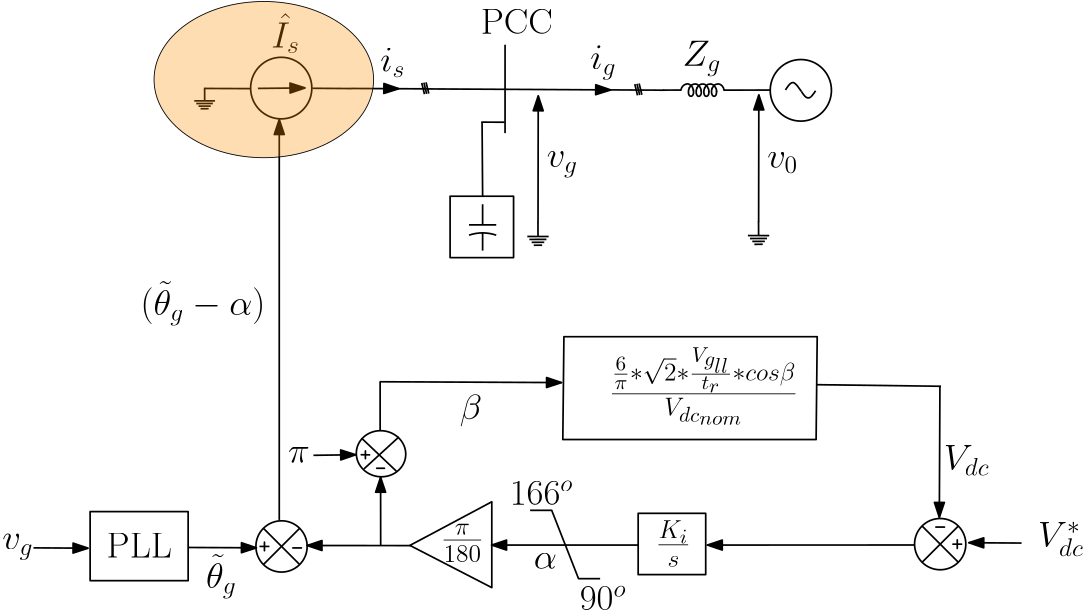}
    \caption{Simplified model-nonlinear}
    \label{fig:simplified}
  \end{subfigure}

  \caption{Models under study}
  \label{fig:Models}
\end{figure}

In the averaged model, hereby referred to as a simplified model-nonlinear, the inverter is represented by a three-phase controlled current source which is modeled to provide the same modulated AC current in fundamental frequency as the actual inverter. The simplified model-nonlinear is shown in Fig. \ref{fig:simplified}.

The vector ${\vec{\boldsymbol{i_s}}} = \hat{I_s}\angle \theta_s$ represents the modulated current vector where $\hat{I_s}$ is the fundamental component of the peak of AC line current and $\theta_s$ is the phase of the AC current. $\hat{I_s}$ and $\theta_s$ can be derived as,
\begin{equation}
 \begin{aligned}
     \hat{I_s} = 4*\frac{\sqrt{3}}{\pi}*I_{dc} *\frac{1}{t_r}, \mbox{ and } \theta_{is}=\tilde\theta_g-\alpha.
\end{aligned}   
\label{eq:current}
\end{equation}

 where $I_{dc}$ is the dc current, $t_r$ is the transformer ratio of the converter transformer, $\theta_g$ is the grid's phase reference from the PLL and $\alpha$ is the firing angle.

 The derivation of $I_s$ follows the well known relationship between AC current and DC current for a six-pulse LCC \cite{b12,b13}. When extended for 12-pulse LCC, the total current is the sum of currents from the two symmetrical six-pulse bridges connected on the secondary side of the converter transformer. The resulting current is then referred to the primary side, yielding the modulated AC current going into the AC grid. As the converter transformer adopted is ideal, the commutation overlap angle is not considered in determining the phase of this current. The phase lag and thus the reactive power consumption is encapsulated in the firing angle $\alpha$ itself. Furthermore, is . The parameters used in the modeling of LCC-HVDC link connected to an AC grid are provided in Table \ref{tab1}.
\begin{table}[h]
\caption{LCC-HVDC and AC grid parameters}
\begin{center}
\scalebox{0.80}{
\begin{tabular}{lll}\toprule
Model component & Parameter & Value \\
\midrule
LCC-HVDC & $V_{dc_{nom}}$ & 500 kV \\
         & $I_{dc_{nom}}$ & 2000 A \\
         & $P_{n,LCC}$ & 1000 MW \\
Converter Transformer (ideal) & $V_{ll,p}$ & 400 kV \\
  & $V_{ll,s}$ & 220 kV \\
Capacitor Bank & $Q_c$ & 600 Mvar \\
AC grid & $U_n$ & 400 kV \\
        & $Z_g$ & variable \\\botrule
\end{tabular}
}
\label{tab1}
\end{center}\vspace*{-18pt}
\end{table}
 
A  constant DC voltage control, \color{black} designed for a response time (settling time of 5\%, $T^{5\%}_r$) of 300ms, \color{black}is employed in the inverter. The controller is implemented as a pure integrator, while the plant is represented by an ideal DC voltage equation. The DC voltage measured is an estimated value of the DC voltage. The DC voltage control and PLL design parameters are provided in Table \ref{tab2}

\begin{table}[h]
\caption{DC voltage control and PLL design parameters}
\begin{center}
\scalebox{0.80}{
\begin{tabular}{lll}\toprule
Model component & Parameter & Value \\
\midrule
DC voltage control & Response Time ($T^{5\%}_r$) & 300ms \\
         & $K_i$ & 1000 deg/pu/s \\
PLL & $\omega_n$ & 50 rad/s \\
  & $\zeta$ & 0.707 \\\botrule
\end{tabular}
}
\label{tab2}
\end{center}\vspace*{-18pt}
\end{table}

\subsection{Simplified linear model of LCC-HVDC link}
 
 The simplified linear model in state-space is developed in dq frame based on the simplified model-nonlinear in Fig. \ref{fig:simplified} using the component connection method \cite{b14}. The two components are LCC-HVDC link and the grid.

The state variables for the network with LCC-HVDC link connected to a thevenin grid are provided below:
\begin{align}
\mathbf{X_{LCC}} = \begin{bmatrix} v_{gx} & v_{gy} & \zeta_{pll} & \Delta\theta_{pll} \end{bmatrix}, \mbox{ }
\mathbf{X_{grid}} = \begin{bmatrix} i_{gx} & i_{gy}  \end{bmatrix}.
\end{align}

where, $v_{gx}$, $v_{gy}$ are the capacitor states, $\zeta_{pll}$, $\Delta\theta_{pll}$ are the PLL states and $i_{gx}$, $i_{gy}$ are the states of the grid inductance. The differential equation governing the dynamics of LCC-HVDC is provided in Appendix A1.

\subsection{Modelling of GFM controlled VSC}

\begin{figure}[h]
     \centering
     \includegraphics[width=1\linewidth]{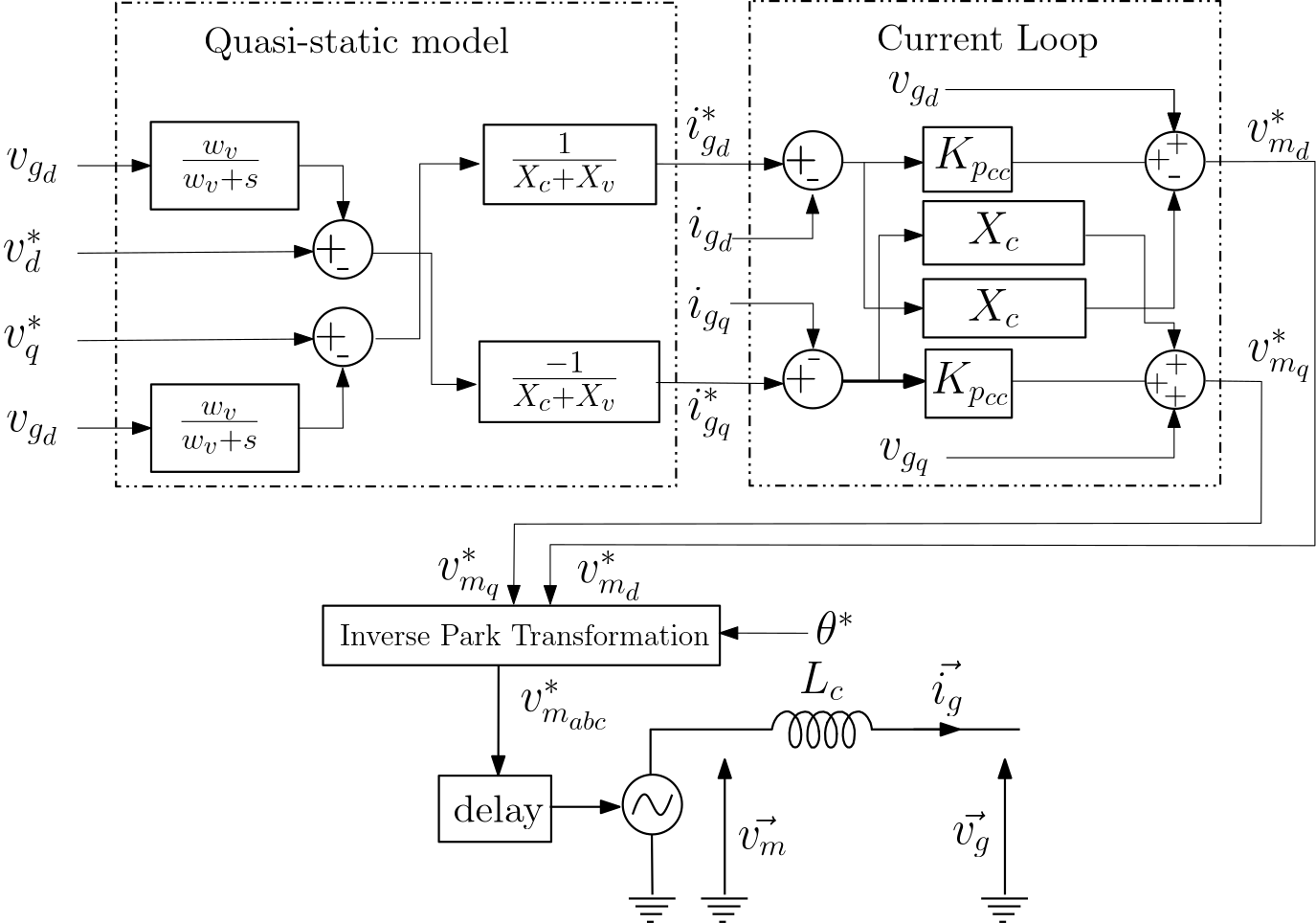}
     \caption{ccGFM configuration}
     \label{fig:CC-GFM}
 \end{figure}
 
\begin{figure}[h]
     \centering
     \includegraphics[width=1\linewidth]{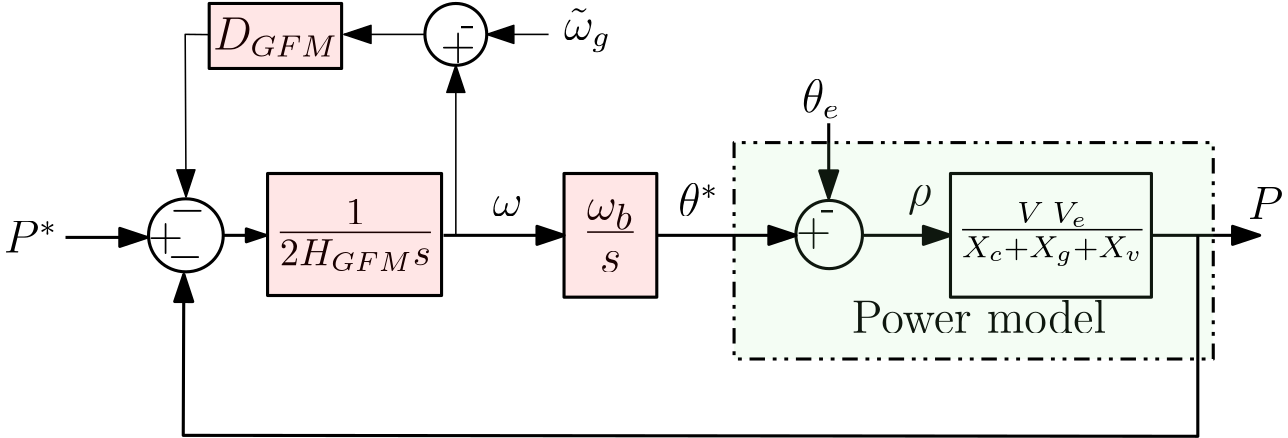}
     \caption{VSM control with PLL}
     \label{fig:VSM}
 \end{figure}

The small-signal state-space model of GFM controlled VSC is based on the model presented in \cite{b11}. The configuration adopted is the one with current control called the current controlled GFM or ccGFM as shown in Fig. \ref{fig:CC-GFM}, hereby referred to as GFM-VSC. \color{black}The ccGFM control approach was chosen for this study because it is widely used for GFM control. \color{black}Virtual synchronous machine (VSM) control with PLL shown in Fig. \ref{fig:VSM} is adopted for power synchronization. The state variables of the GFM-VSC are provided below:
\begin{equation}
\mathbf{X_{GFM}} = \begin{bmatrix} i_{sx}, i_{sy}, \zeta^\theta _m, \theta_{m}, v_d^f, v_q^f, \zeta^{i_d}, \zeta^{i_q},\\ P_x, P_y, \zeta^{PLL}, \tilde{\theta_g}, v^f_{g_x}, v^f_{g_y}, i^f_{s_x}, i^f_{s_y} \end{bmatrix}   
\end{equation}

The system and control parameters of the GFM-VSC model are provided in Table \ref{tab3} and Table \ref{tab4}, respectively.
\begin{table}[h]
\caption{GFM-VSC system parameters}
\begin{center}
\scalebox{0.80}{
\begin{tabular}{lll}\toprule
Model component & Parameter & Value \\
\midrule
GFM & $S_{n_{GFM}}$ & 1000 MVA \\
         &  $R_c$ / $L_c$  &  0.005 pu / 0.15 pu  \\
OHL &  $R_{23}$ / $L_{23}$ & 0.0072 pu / 0.144 pu \\
\botrule        
\end{tabular}
}
\label{tab3}
\end{center} 
\caption{GFM-VSC control parameters}
\begin{center}
\scalebox{0.80}{
\begin{tabular}{lll}\toprule
Model component & Parameter & Value \\
\midrule
VSM control & $D_{GFM} / H_{GFM}$  &    118.15/ 5 s \\
         &  & ($\zeta=1$) \\
Virtual Impedance & $X_v$ & 0.3 pu \\
 Voltage filter  & $\tau_{f_{CC}}$ & 40ms \\
Current controller     &  $\omega_{CC}$ & 1200 rad/s \\ \botrule 
\end{tabular}
}
\label{tab4}
\end{center}\vspace*{-18pt}
\end{table}

\section{Validation of the models}

\subsection{Validation of the LCC-HVDC model}
\begin{figure}[h!]
     \centering
     \includegraphics[width=1\linewidth]{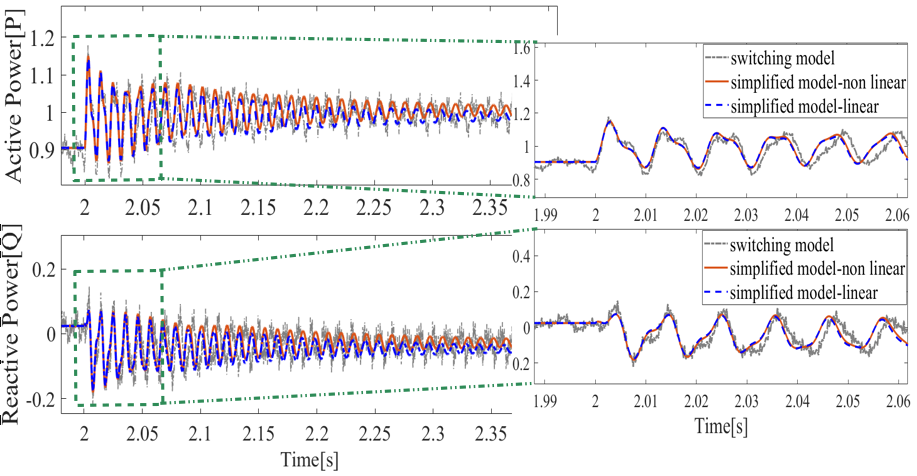}
     \caption{Model Validation: PQ response}
     \label{fig:LCC_thev_validation}
 \end{figure}

The fidelity of the linearized small-signal model is verified by validating the results against the EMT simplified non-linear model and EMT switching model. The active and reactive power response of the three models have been compared for a step \color{black}of 0.1 pu \color{black} in DC current at an SCR of 5. \color{black} DC current perturbation propagates through the DC link and excites the dominant AC-DC dynamics at the inverter AC terminals. \color{black} The SCR is obtained looking into the thevenin grid from the PCC (Ref \ref{fig:switch}) and it is the inverse of the per-unit grid impedance, $Z_g$, for a thevenin equivalent.  The validation results in Fig. \ref{fig:LCC_thev_validation} show a good agreement. Hence, the small signal model can be used for further stability study.



\FloatBarrier

\subsection{Validation of the benchmark system with GFM-VSC}

\begin{figure}[h]
    \centering
    \includegraphics[width=1\linewidth]{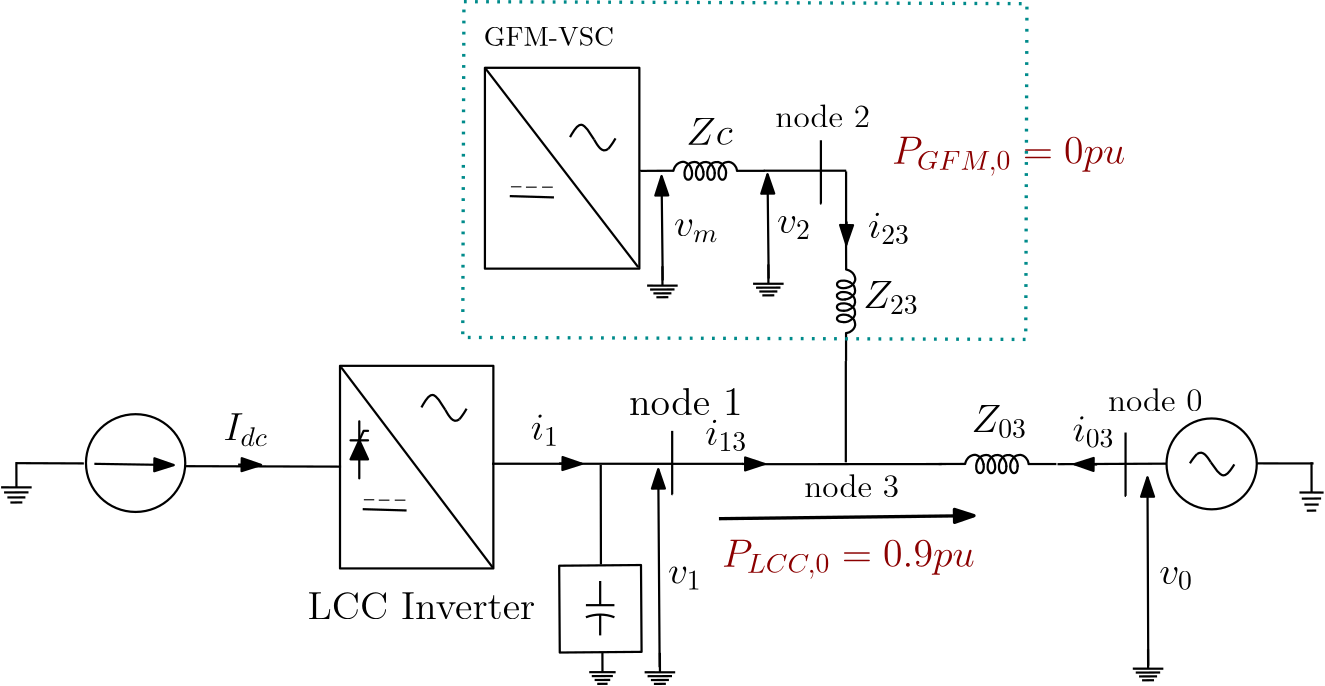}
    \caption{Setup with GFM-VSC, LCC-HVDC and Infinite grid}
    \label{fig:benchmark setup}
\end{figure}

A GFM-VSC was connected to the PCC through a 30 km over-head line (OHL) in the previous setup to study its impact on the stability of the LCC-HVDC. The new setup is shown in Fig. \ref{fig:benchmark setup}. 

The validation of the system is performed at an SCR of 3 with a step on the DC current as the event. The validation shows a good agreement.
\begin{figure}[h]
    \centering
    \includegraphics[width=1\linewidth]{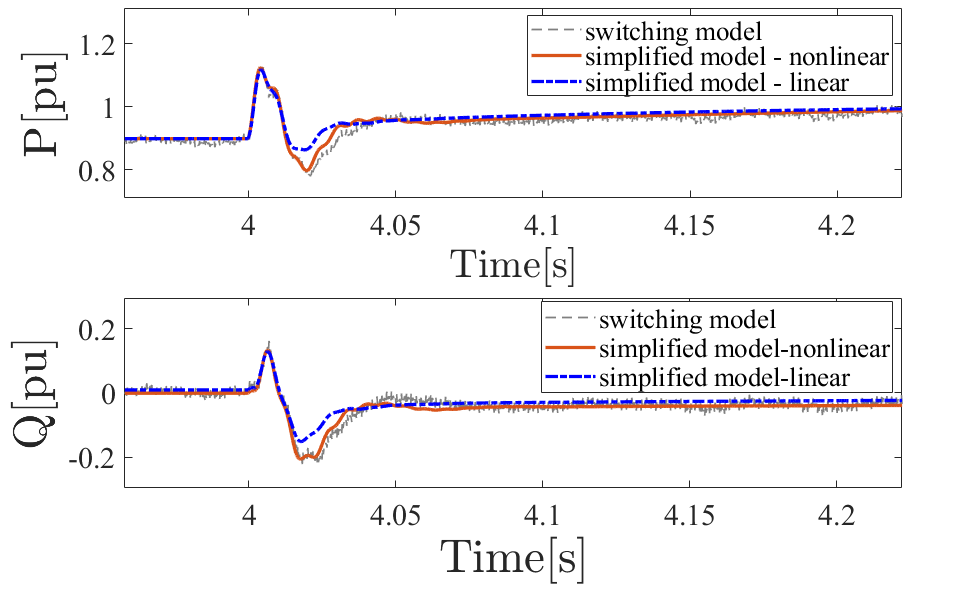}
    \caption{Model validation with GFM-VSC}
    \label{fig:Validation2}
\end{figure}

\section{Results}
First, the LCC-HVDC link connected to an infinite grid, without GFM-VSC, is considered. The system adopted is the one in Fig. \ref{fig:benchmark setup} without the part bounded by the dotted rectangle.
\subsection{Small signal stability analysis of LCC-HVDC link connected to a weak grid}
\begin{figure}[h]  
  \centering
    \includegraphics[width=1\linewidth]{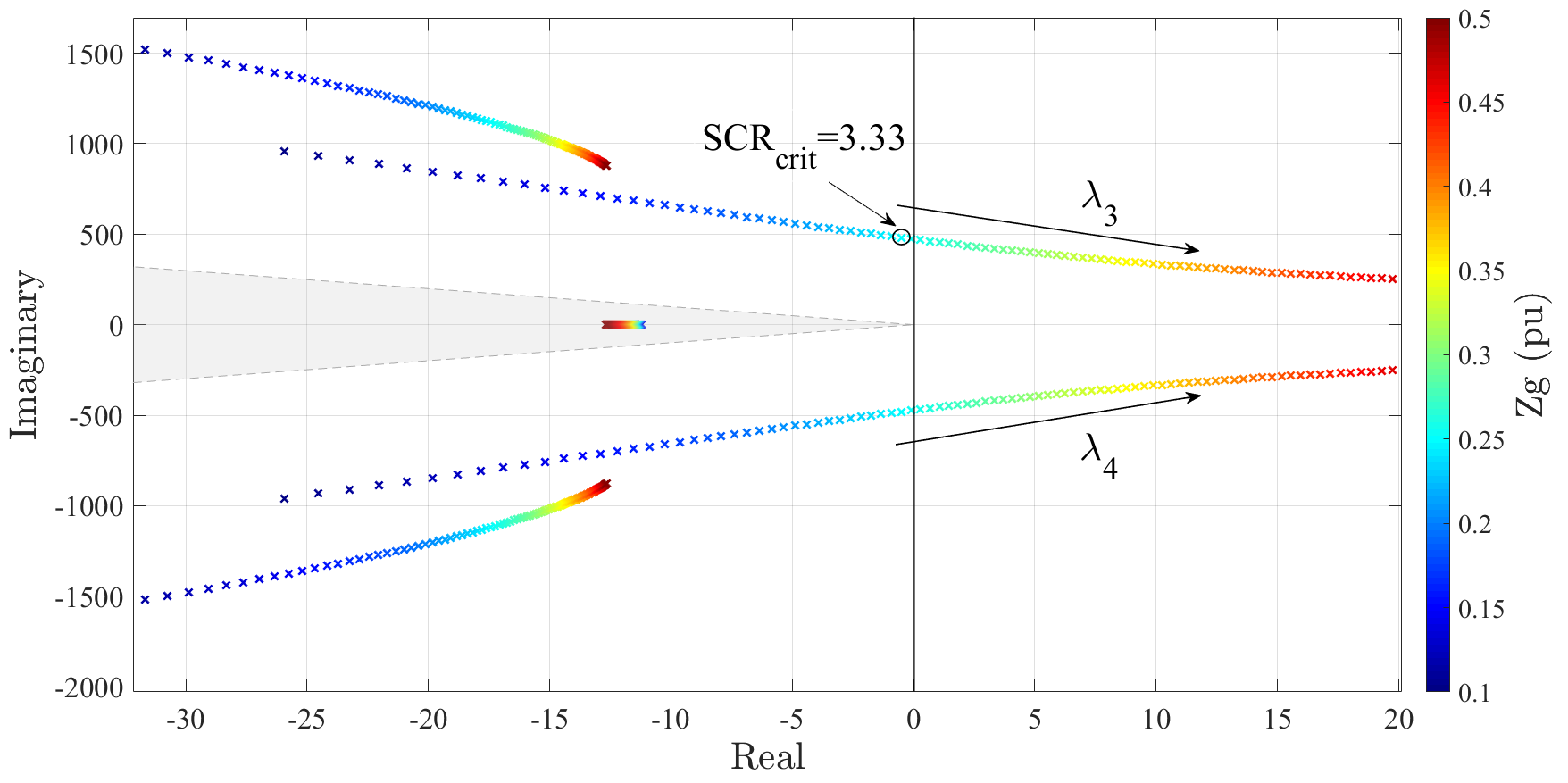}
    \caption{Sensitivity to grid strength variation}
    \label{fig:Sensitivity_Lg}
\end{figure}

The sensitivity to grid strength is studied by increasing the grid impedance, $Z_g$ (Ref Fig. \ref{fig:switch}). In a thevenin equivalent setup, the grid impedance in per-unit and SCR have an inverse relation. Thus, increasing the grid impedance reduces the SCR. From the sensitivity analysis, the stability limit is found to be at SCR of 3.33 as shown in Fig. \ref{fig:Sensitivity_Lg}. The eigenvalue pair $\lambda_{3,4}$ (-0.51 $\pm$ j481,$\zeta$ = 0,1\%, $f_{osc}$=76.5 Hz) is the most critical mode. 

\begin{figure}[h]  
    \centering
    \includegraphics[width=1\linewidth]{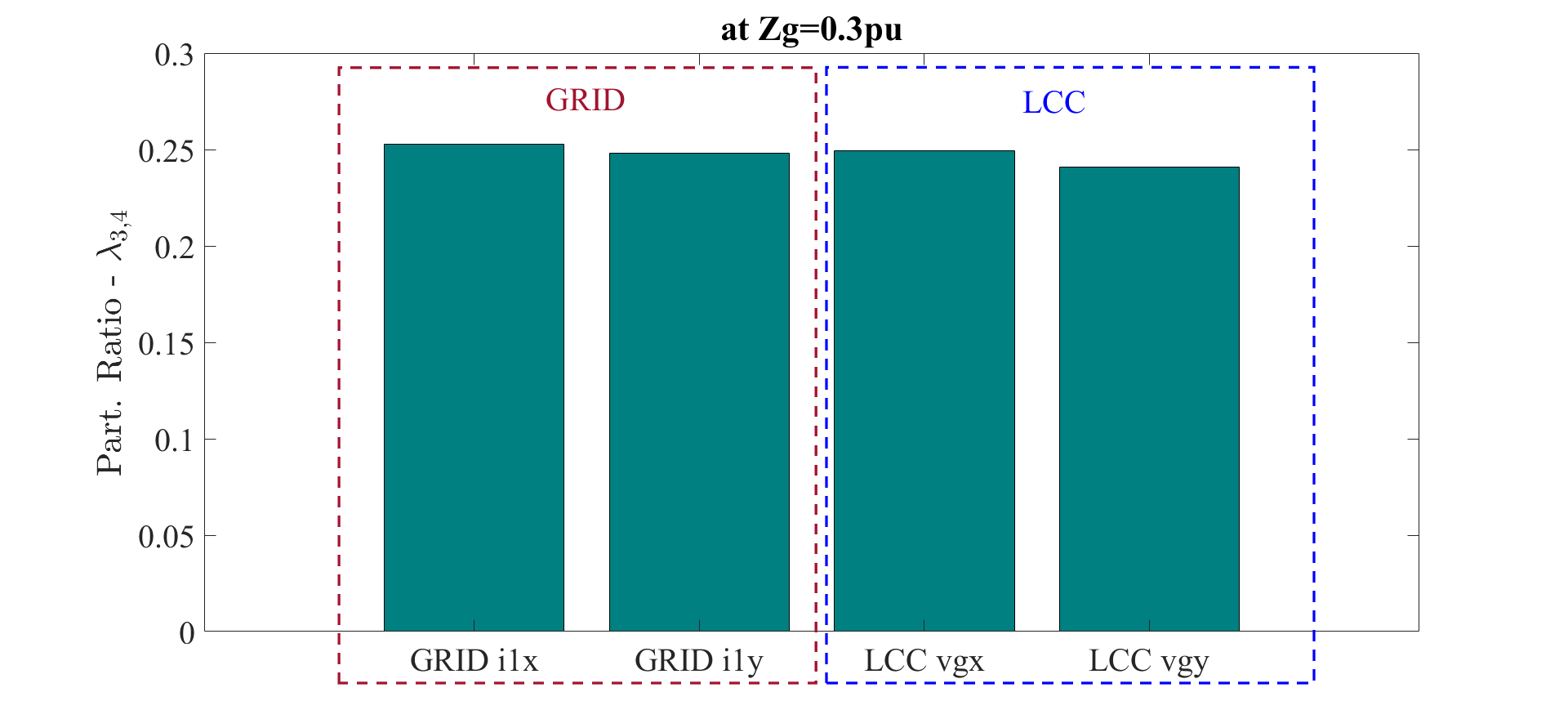}
    \caption{PF: LCC-HVDC connected to thevenin equivalent}
    \label{fig:PF_LCC}
\end{figure}

The participation factor (PF) can provide information about which state variables are responsible for a mode's behavior \cite{b15a,b15,b16} and the PF analysis of the critical mode in Fig. \ref{fig:PF_LCC} showed the participation of both the LCC and grid states.

\subsection{Stability enhancement with the addition of a GFM VSC}

\begin{figure}[h]
    \centering
    \includegraphics[width=1\linewidth]{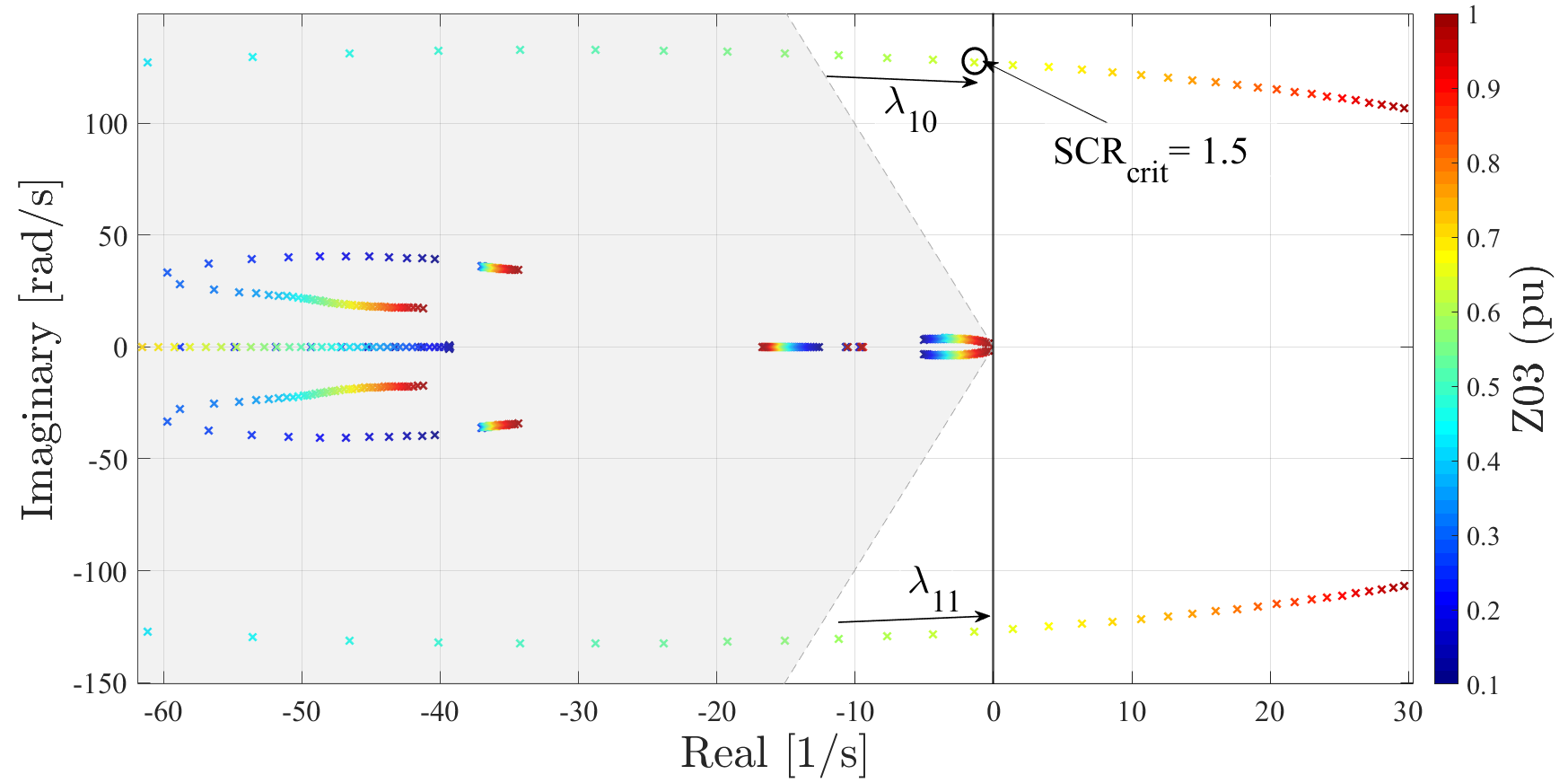}
    \caption{Sensitivity to grid strength with GFM-VSC}
    \label{fig:Sensitivity_CC}
\end{figure}

A 1000 MVA GFM power converter is included now, as shown in Fig. \ref{fig:benchmark setup}. The power flow is set up so that there is no active power injection from the GFM-VSC: $P_{GFM,0}=0$. Thus, the \color{black}generic GFM converter \color{black}could represent an enhanced STATCOM (E-STATCOM) or \color{black} an energy-storage system (ESS) interfaced by a power converter, such as batteries (BESS) or ultracapacitors (UCAP). \color{black}  The sensitivity analysis to grid strength in Fig. \ref{fig:Sensitivity_CC} showed that the stability limit improved from SCR=3.33 to SCR=1.5 with the addition of GFM-VSC.

\begin{figure}[h]
    \centering
    \includegraphics[width=1\linewidth]{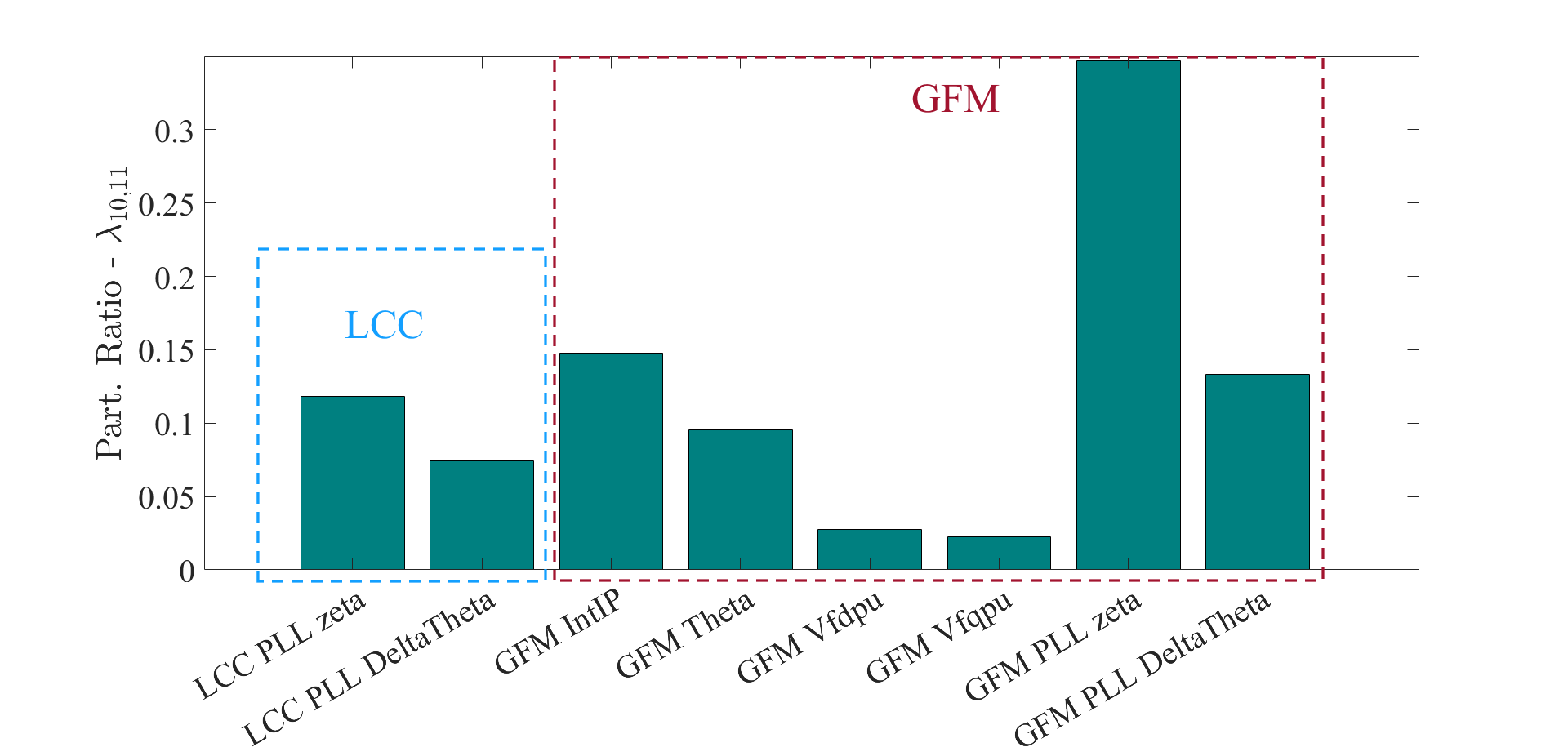}
    \caption{PF with GFM VSC connected}
    \label{fig:PF CCGFM}
\end{figure}

There are two dominant modes $\lambda_{10,11}$ and $\lambda_{20,21}$ among which  $\lambda_{10,11}$(-1.39 $\pm$ j127.22, $\zeta$=1.1\%, $f_{osc}$=20.25 Hz) is the most sensitive to the decrease in SCR. Additionally, the analysis of participation factor of the critical mode ($\lambda_{10,11}$) in Fig. \ref{fig:PF CCGFM} showed a much higher participation of the states of GFM-VSC. This provides an important conclusion: The contribution
of LCC states in the critical mode decreases significantly with the addition of GFM-VSC, while the absence of grid's states show the decoupling of LCC-HVDC from weak grid and the dynamics is governed by the GFM-VSC's control.

\subsection{Sizing a GFM power converter to ensure small signal stability}
A study was performed to determine the minimum GFM-VSC capacity required to maintain the stability of the system in a weak grid scenario. To emulate a weak grid, the SCR of the grid was set to SCR=1.8 and a sensitivity analysis was performed with respect to the GFM VSC capacity.

\begin{figure}[h]
    \centering
    \includegraphics[width=1\linewidth]{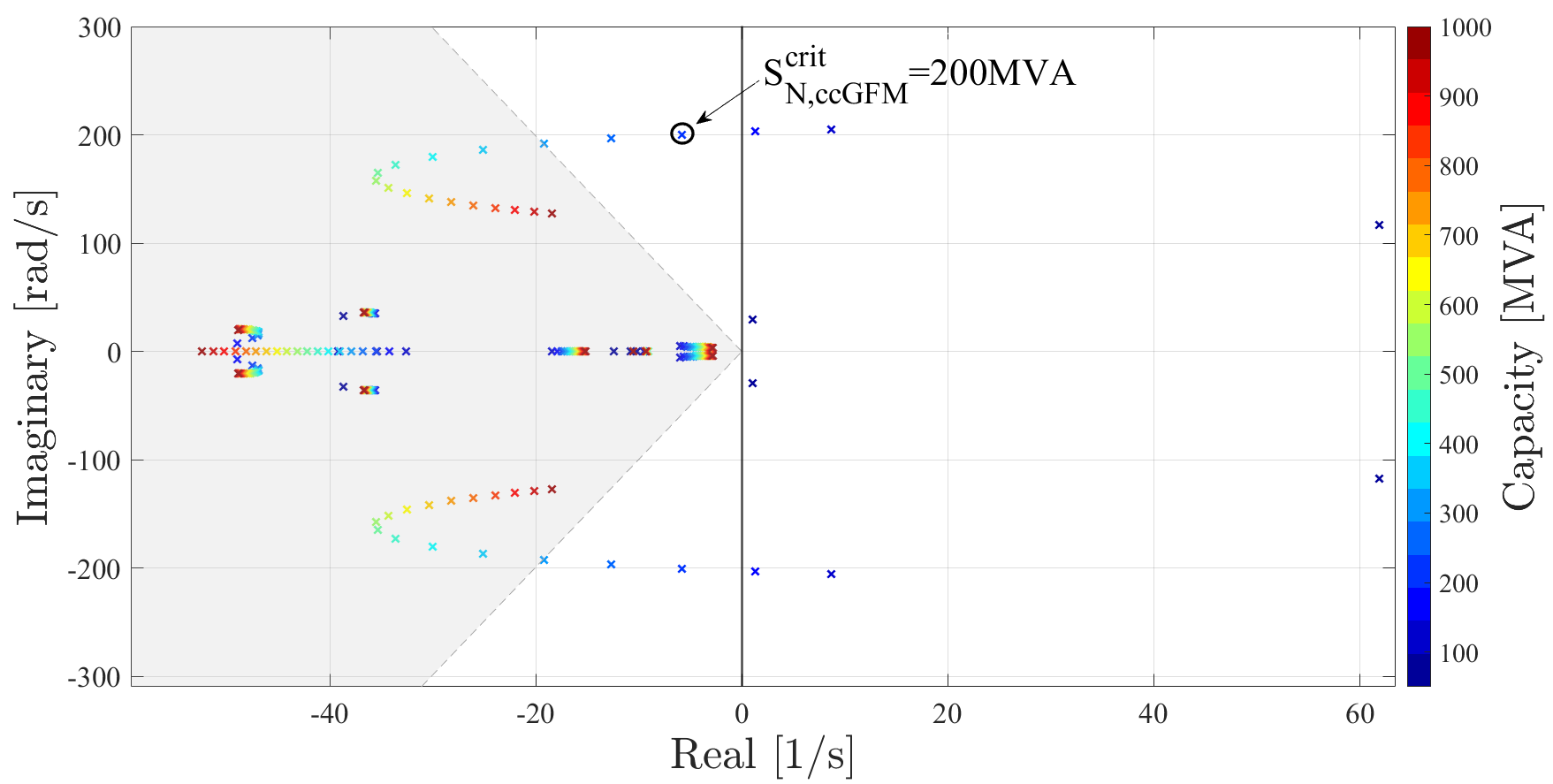}
    \caption{Sensitivity to GFM VSC capacity}
    \label{fig:Capacity_CC}
\end{figure}
\begin{figure}[h]
    \centering
     \includegraphics[width= 1\linewidth]{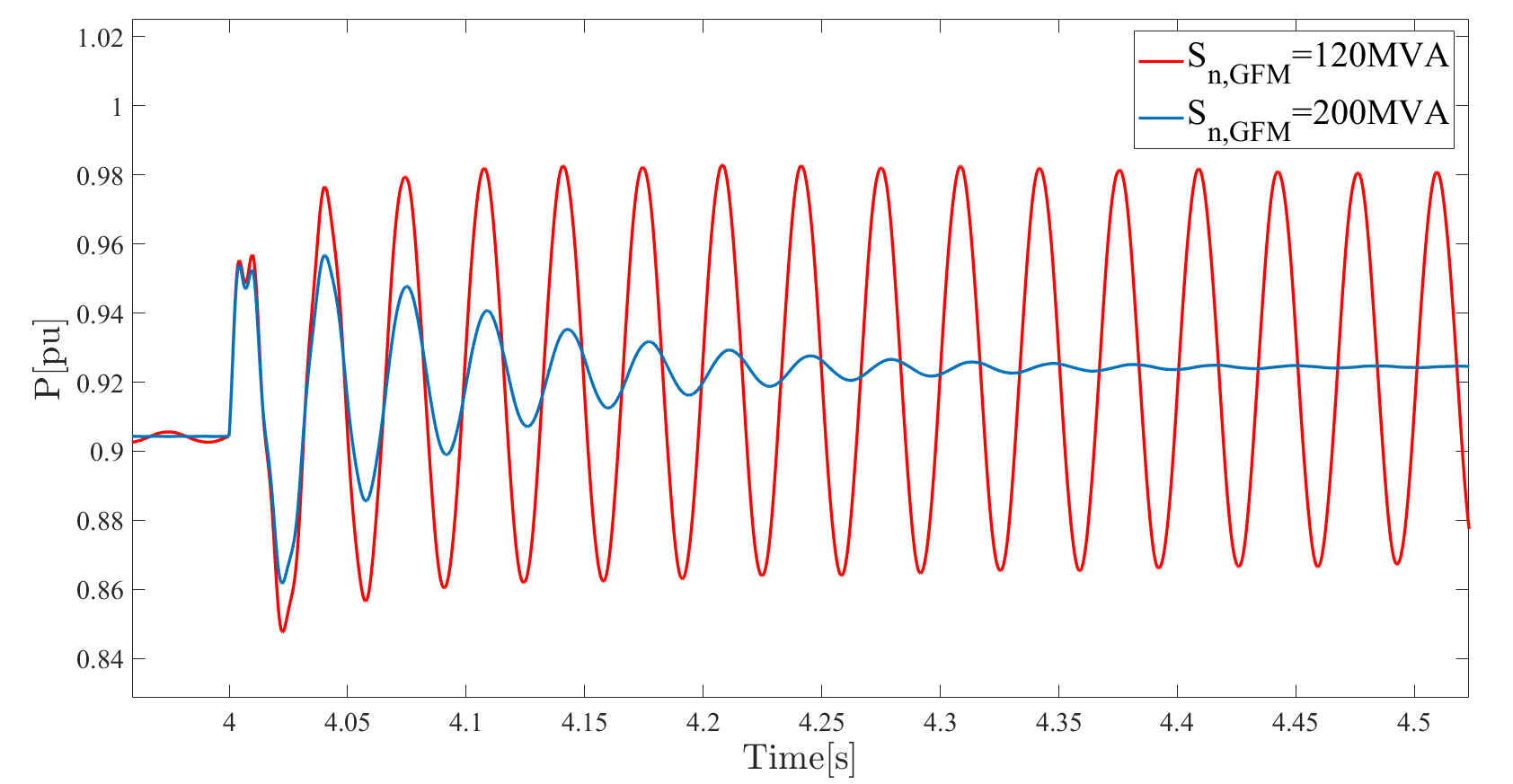}  
    \caption{Simulation verification of ccGFM capacity for stability}
    \label{fig:ccGFM cap verification}
\end{figure}

The result in Fig. \ref{fig:Capacity_CC} indicates that, for the nominal LCC-HVDC capacity of 1000 MW, the minimum amount of GFM-VSC apparent power required to maintain stability was 200 MVA, which corresponds to 16.7$\%$ of the total nominal apparent power. For a target damping ratio of 10\%, the required capacity increases to 360 MVA or 26$\%$ of the total nominal apparent power. 
The findings of the sensitivity analysis are further reaffirmed through MATLAB/Simulink simulation. Two cases with a GFM-VSC capacity of 120 MVA and 200 MVA are simulated and stability is analyzed from the AC side active power behavior for a small step in DC current . From the results in Fig. \ref{fig:ccGFM cap verification} we can clearly see that the system is unstable for GFM-VSC capacity of 120 MVA ($f_{osc}$ = 30 Hz) and remains stable for GFM-VSC capacity of 200 MVA.

\section{Conclusion}
This paper investigated the stabilizing influence of the GFM-VSC power converter on an LCC-HVDC link connected to a weak grid, in terms of small-signal stability and interactions. The analysis showed that \color{black}with a GFM-VSC power converter,  \color{black} the stability limit of the LCC-HVDC link improves from an SCR of 3.33 to 1.5. Furthermore, the study on sizing of GFM-VSC revealed that the minimum amount of GFM-VSC capacity required for stability to be 16.7\% of the total nominal apparent power, for the test system analyzed in this work. It is important to mention that the sizing of the GFM-VSC power converter to ensure the stability of the LCC-HVDC depends on many aspects: test system considered, control system of the LCC-HVDC link and the GFM-VSC, and location of the latter. Furthermore, selection of the nominal apparent power of the GFM-VSC will also depend on the overall stability of the system, which include relevant phenomena, such as small-signal stability, commutation failure or voltage stability, among others. This paper only focuses on small-signal stability and interactions and it concludes that a GFM-VSC can stabilize the system in terms of small-signal stability and \color{black} the paper \color{black} provides guidelines to size it, but numerical results are related to each particular system and should not be generalized. This study was carried out \color{black} using \color{black}  a simplified model of LCC-HVDC and including the rectifier-side dynamics to have the complete LCC-HVDC representation could provide additional insights into the system interaction. Thus, future work will therefore extend the analysis with a complete LCC-HVDC model to further validate the conclusions and assess the accuracy and relevance of the simplified model.

\section{Acknowledgements}
This paper has received funding from the European Union’s Horizon Europe Research and Innovation programme under the Marie Skłodowska-Curie grant agreement No. 101073554. The authors would like to thank Dr. Edgar Nuño Martínez and Mr. Antonio Cordón (Red Eléctrica - Redeia) for their useful feedback and comments.

\section{References}

\section*{APPENDIX A}
\renewcommand{\theequation}{A\arabic{equation}}
\setcounter{equation}{0}  
The differential equations governing the dynamics of LCC are provided below:
\begin{equation}
\label{eq:LCC}
{
\begin{aligned}
\dot v_{gx} &= \frac{i_{sx}}{C_1} - \frac{i_{gx}}{C_1} + \omega_g * v_{gy} \\
\dot v_{gy} &= \frac{i_{sy}}{C_1} - \frac{i_{gy}}{C_1} - \omega_g * v_{gx}\\
\dot i_{gx} &= \frac{v_{gx}}{L_g} - \frac{v_{sx}}{L_g} - \frac{R_g}{L_g} *i_{gx} + \omega_g * i_{gy}\\
\dot i_{gy} &= \frac{v_{gy}}{L_g} - \frac{v_{sy}}{L_g} - \frac{R_g}{L_g} *i_{gy} - \omega_g * i_{gx}\\   
\dot \zeta_{pll} &= K_{i_{pll}} * v_{gq_{pll}} \\
\dot \Delta\Theta_{pll} &= \omega_b*(\Delta\omega_{pll})
\end{aligned}
}
\end{equation}
DC voltage equation and its linearization:
\begin{equation}
\label{eq:DC Voltage}
{
\begin{aligned}
V_{dc} &= -\frac{6*\sqrt2}{\pi} * \frac{1}{t_r}*V_{g_{pcc}}*cos\alpha \\
V_{dc} &=  -\frac{6*\sqrt2}{\pi}* \sqrt{v^2_{gd_{pll}}+v^2_{gq_{pll}}} *cos\alpha \\
\Delta V_{dc} &= k_{vd}*\Delta v_{gd_{pll}} + k_{vq}*\Delta v_{gq_{pll} + k_\alpha *\Delta\alpha} \\
k_{vd} &= \frac{\partial V_{dc}}{\partial v_{gd}}, k_{vq} = \frac{\partial V_{dc}}{\partial v_{gq}}, k_\alpha = \frac{\partial V_{dc}}{\partial \alpha}
\end{aligned}
}
\end{equation}

The differential equations governing the dynamics of ccGFM are provided below:
\begin{equation}
\label{eq:ccGFM}
{
\begin{aligned}
\dot i_{sx} &= \frac{\omega_b}{L_c} (v_{mx} - v_{gx} -R_c *i_{gx} + \omega_g*L_c*i_{gy}) \\
\dot i_{sy} &= \frac{\omega_b}{L_c} (v_{my} - v_{gy} -R_c *i_{gy} - \omega_g*L_c*i_{gx}) \\
\dot v^f_d &= \omega^{QSEM}_{LPF}(v_{gd}-v^f_d) \\
\dot v^f_q &= \omega^{QSEM}_{LPF}(v_{gq}-v^f_q) \\
\dot \zeta_{id} &= K^{CC}_i (i^*_d - i_{gd}) \\
\dot \zeta_{iq} &= K^{CC}_i (i^*_q - i_{gq}) \\
\dot \zeta &= \frac{1}{2H}(P^*-P + k_d*(\tilde{\omega_g}-\omega)\\
\dot \Theta_{m} &= \omega_b*(\zeta-w^*)
\end{aligned}
}
\end{equation}

Note: In the above equations, subscript x and y denote the grid's reference frame, and the subscript d and q denote the converter's frame. $\omega_g$ is the grid frequency in pu, $\omega_b$ is the base frequency in rad/s and $\omega^*$ is the frequency setpoint set to 1.

\end{document}